\documentclass[aps,prl,preprintnumbers,english,footinbib,floatfix,reprint,twocolumn,showpacs,amsmath,amssymb,superscriptaddress,longbibliography,10pt,tightenlines]{revtex4-2}
\pdfoutput=1

\usepackage[utf8]{inputenc}
\usepackage{natbib}
\usepackage{multirow}
\usepackage{physics}
\usepackage{amsmath,amssymb,amsthm}
\newtheorem{theorem}{Theorem}[section]
\usepackage{subcaption}
\usepackage{natbib}
\usepackage{cancel}
\usepackage{comment}
\usepackage[english]{babel}
\usepackage[pdftex]{graphicx}
\usepackage[lflt]{floatflt}
\usepackage{epstopdf}
\usepackage[normalem]{ulem}
\usepackage{subcaption}
\usepackage{ragged2e}
\DeclareCaptionJustification{justified}{\justifying}
\usepackage[final]{hyperref} 
\hypersetup{
	colorlinks=true,       
	linkcolor=red,        
	citecolor=red,        
	filecolor=magenta,     
	urlcolor=blue         
}



\usepackage{color}
\usepackage[dvipsnames]{xcolor}

\begin{document}

\title{Universal Phase Locking from Slowly Varying Inhomogeneity in Tunnel-Coupled One-Dimensional Condensates}

\author{Bence Fitos}\email{fitosbence@edu.bme.hu}
\affiliation{Department of Theoretical Physics, Institute of Physics, Budapest University of Technology and Economics, H-1111 Budapest, M{\H u}egyetem rkp.~3.}
\affiliation{HUN-REN-BME-BCE Quantum Technology Research Group, Institute of Physics, Budapest University of Technology and Economics, M{\H u}egyetem rkp.~3., H-1111 Budapest, Hungary}
\author{G\'abor Tak\'acs}\email{takacs.gabor@ttk.bme.hu}
\affiliation{Department of Theoretical Physics, Institute of Physics, Budapest University of Technology and Economics, H-1111 Budapest, M{\H u}egyetem rkp.~3.}
\affiliation{BME-MTA Momentum Statistical Field Theory Research Group, Institute of Physics, Budapest University of Technology and Economics, M{\H u}egyetem rkp.~3., H-1111 Budapest, Hungary}
\date{November 1, 2026}

\begin{abstract}
The sine-Gordon model can be experimentally realised by tunnel-coupled one-dimensional Bose condensates. However, the experiments show relaxation to a phase-locked state, which conflicts with the predictions of the homogeneous integrable theory. Here we combine the truncated Wigner approximation with an asymptotic analysis, and show that this locking is produced by a slowly varying density profile through a universal mechanism, whereby the relative phase near the density maximum drifts towards zero. We show that this drift is already present in the linearised theory and its timescale is determined by the density curvature alone, while it is
independent of the tunnel coupling. Rescaling time by the relaxation time scale
collapses the dynamics for different system sizes and density profiles, while
non-linearity suppresses the excitation returning from the edges.
\end{abstract}

\maketitle

\textit{Introduction.}--- 
The non-equilibrium dynamics of isolated quantum many-body systems remains a central challenge in modern physics \cite{2011RvMP...83..863P, 2016RPPh...79e6001G}. Tunnel-coupled one-dimensional Bose gases provide a particularly attractive setting for studying such dynamics, as their low-energy behaviour is described by the quantum sine-Gordon model \cite{2020ScPP....9...25V, 2021ScPP...10...90V}, an experimentally realisable interacting quantum field theory \cite{2010PhRvL.105s0403C}. In the homogeneous limit, this model belongs to a special class of integrable field theories whose exact spectrum and scattering properties are completely known \cite{1979AnPhy.120..253Z, 1995IJMPA..10.1125Z}. A prominent feature of post-quench dynamics in this system is universal dephasing: following a sudden change in the tunnel coupling, coherent Josephson oscillations decay according to universal scaling laws that are largely independent of microscopic details \cite{2013PhRvL.110i0404D, 2019PhRvA.100a3613H}. Experimental observations have established this phenomenon as a hallmark of non-equilibrium dynamics in low-dimensional quantum gases \cite{2007Natur.449..324H, 2012Sci...337.1318G}.

Real experiments, however, are intrinsically inhomogeneous. The trapping potential generates a spatially varying density profile, thereby breaking the integrability of the homogeneous 1D Bose gas \cite{2020PhRvL.125x0604B}. While often regarded as a weak perturbation, such inhomogeneity can qualitatively alter the long-time dynamics \cite{2019PhRvL.123m0602B,2022ScPP...12..144H}. It may represent the dominant distinction between experimentally realised condensates and the idealised integrable theory. Indeed, the long-time dynamics of coupled condensates is significantly altered under these conditions; instead of the universal rephasing predicted for homogeneous systems \cite{2013PhRvL.110i0404D}, a robust phase-locking behaviour is experimentally observed \cite{2018PhRvL.120q3601P}. Subsequent three-dimensional stochastic classical-field simulations reproduced this relaxation, attributing the early damping to linear dephasing caused by the density dependence of the Josephson frequency and the persistent locking to non-linear coupling beyond the sine-Gordon description \cite{2021PhRvR...3b3197M}; more recently, effective one-dimensional sine-Gordon--Luttinger simulations showed that phase locking is already recovered once the trap-induced inhomogeneous background is included \cite{2025arXiv251213901F}. 

While these works establish the essential role of the trap, they leave open the question of what sets the locking timescale. In this Letter, we show that slowly varying
inhomogeneity, i.e., a density profile whose curvature is small on the scale
of the correlation length set by the tunnel coupling, produces a universal
phase-locking drift with a timescale that is determined by the curvature. Studying the post-quench dynamics of an inhomogeneous sine-Gordon model in the experimentally relevant semiclassical regime \cite{2016PhRvE..93f2101K, Szasz-Schagrin2022}, we uncover a crossover between two different forms of universal behaviour. At short times, the system follows the universal dephasing predicted by the homogeneous model. In contrast, at parametrically later times it enters a phase-locking regime absent in the homogeneous limit, which is characterised by a systematic drift of the relative phase toward zero \cite{2018PhRvL.120q3601P}. The characteristic timescale of this regime is set by the curvature of the density profile at the trap centre and is independent of the tunnel coupling; the dynamics exhibits a universal scaling collapse which is encoded in the linearised modes of the trapped system.

\begin{figure*}[t]
    \centering
    \subfloat[Universal dephasing\label{fig:dephasing}]{%
    \includegraphics[width=0.50\linewidth]{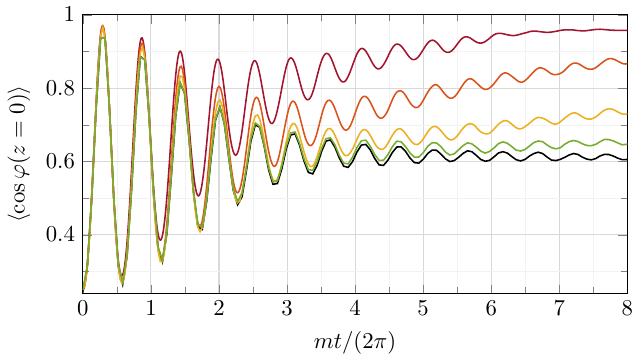}
    }
    \subfloat[Phase locking\label{fig:phaselocking}]{%
    \includegraphics[width=0.50\linewidth]{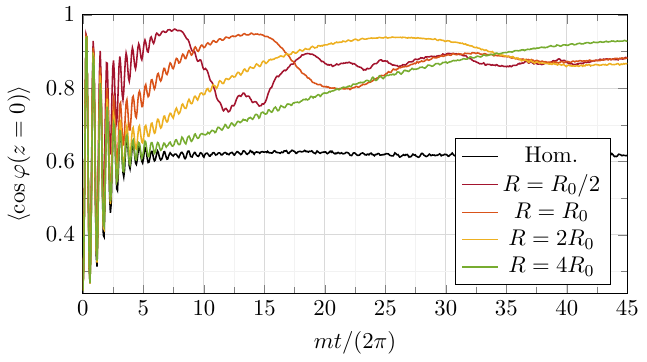}
    }
    \caption{Demonstration of the universal dephasing (a) and phase locking (b) of the (inhomogeneous) sine-Gordon model. The Figure shows the cosine of the relative phase $\varphi$ at the centre of the condensate ($z = 0$) as a function of time (measured in units of Josephson oscillations) for different system sizes ($R/R_0 = 0.5, 1, 2, 4$). The black curve shows the result for the homogeneous profile, $f(z)\equiv 1$. The parameters are chosen to mirror those of the experiment: $R_0 m = 25.5, \beta = 0.343, \langle\varphi_0\rangle = -1.3, \sigma_0=0.2$.}
    \label{fig:dephasing_and_phaselocking}
\end{figure*}

\begin{figure} [t]
    \centering
    \includegraphics[width=1.0\linewidth]{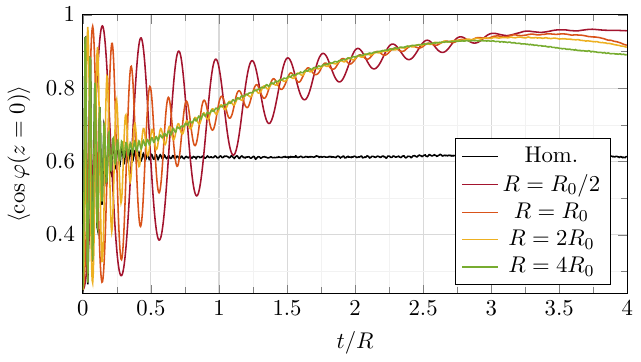}
    \caption{Scaling collapse of phase locking. The Figure shows the cosine of the relative phase $\varphi$ at the centre of the condensate ($z = 0$) as a function of $t/R$ with different system sizes. The black curve shows the result for the homogeneous case. Parameters: $R_0 m = 25.5, \beta = 0.343, \langle\varphi_0\rangle = -1.3, \sigma_0=0.2$.  We observe that it takes approximately $R m/2$ Josephson oscillations to reach the peak value.}
    \label{fig:phaselockingScaled}
\end{figure}

\begin{figure}[t]
    \centering
    \includegraphics[width=1.0\linewidth]{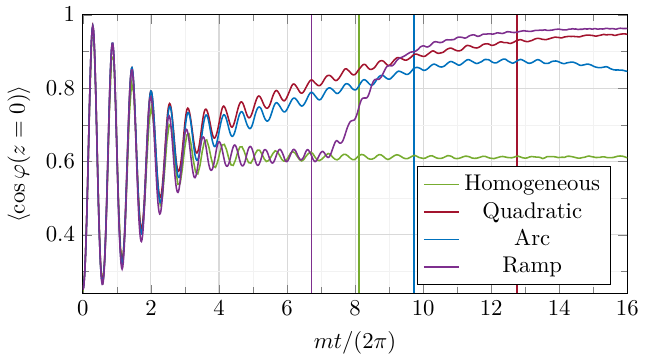}
    \caption{The cosine of the relative phase at the centre of the condensate as a function of time (measured in units of Josephson oscillations) with different profiles $f(z)$. Vertical lines mark the light-cone time for each profile (cf.~Eq.~\eqref{eq:lightconetime}). Parameters: $R m = 25.5, \beta = 0.343, \langle\varphi_0\rangle = -1.3, \sigma_0=0.2$.} \label{fig:mixed}
\end{figure}

\begin{figure}[t]
    \centering
    \includegraphics[width=1.0\linewidth]{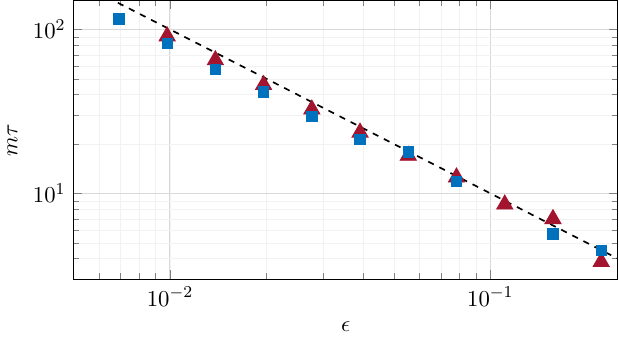}
    \caption{Fitted relaxation time $\tau$ [Eq.~(S.2)] as a function of the
    curvature parameter $\epsilon=\sqrt{-\partial_x^2 f(0)/2}\,/m$, for the
    quadratic (triangles, $f(z)=1-z^2$) and arc (squares, $f(z)=\sqrt{1-z^2}$)
   profiles. The dashed line is the parameter-free prediction of Eq.~(2),
   $m\tau_{\rm rel}=1/\epsilon$. Fitted exponents (std.\ error): $-0.974(22)$
   (quadratic) and $-0.936(17)$ (arc).}
    \label{fig:loglogtau}
\end{figure}

\begin{figure} [t]
    \centering
    \includegraphics[width=1.0\linewidth]{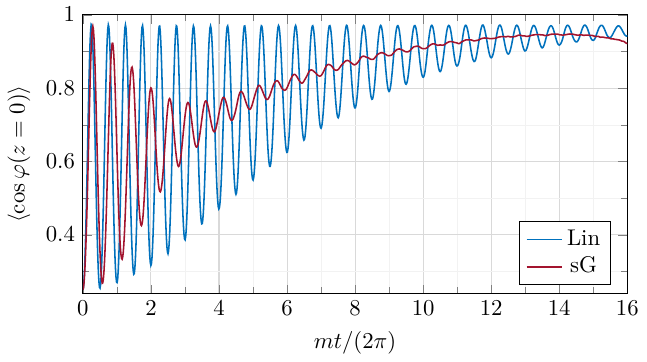}
    \caption{The sine--Gordon vs.~the linearized model. The Figure shows the cosine of the relative phase $\varphi$ at the centre of the condensate ($z = 0$) as a function of time for the sine--Gordon model and for the corresponding linearised theory. Parameters:  $R m = 25.5, \beta = 0.343, \langle\varphi_0\rangle = -1.3, \sigma_0=0.2$.}
    \label{fig:vslin}
\end{figure}

\textit{Model and quench protocol.}--- 
Tunnel-coupled one-dimensional Bose condensates realise sine-Gordon dynamics \cite{2020ScPP....9...25V, 2021ScPP...10...90V}; in particular, they provide a controlled setting to investigate the effects of inhomogeneity induced by a slowly varying trap on non-equilibrium dynamics \cite{2022ScPP...12..144H}. Motivated by recent experiments \cite{2017Natur.545..323S, 2018PhRvL.120q3601P}, we consider an inhomogeneous sine-Gordon model with action
\begin{align} \label{eq:actionDim}S = \int dt\int_{-R}^R dx \Big(&\frac12 (\partial_t\phi)^2-\frac12 f(x/R)(\partial_x\phi)^2\nonumber\\
&+\frac{m^2}{\beta^2}f(x/R)\cos(\beta\phi)\Big)\,,\end{align}   
where $f(z) = 1 - z^2$ describes the density profile of a harmonically trapped condensate \cite{2005EPJD...35..147B, 2021PhRvR...3b3197M}, with $f(z)=1$ corresponding to the homogeneous limit. We introduce the rescaled field $\varphi=\beta\phi$. The dynamics is controlled by two dimensionless parameters: $\beta$ sets the strength of quantum fluctuations, while $mR$ parametrises the system size in units of the correlation length at the centre. Slowly varying inhomogeneity corresponds to a density profile whose curvature at the centre is small on the scale of the correlation length $1/m$: $\epsilon\equiv\sqrt{-\partial_x^2 f(0)/2}\,/m\ll 1$. For the harmonic trap $f(z)=1-z^2$ used in the experiments, $\epsilon=1/(mR)$, so slow variation is precisely the condition $mR\gg 1$. We note that the spatial dependence of $f(z)$ breaks integrability of the homogeneous sine-Gordon model. 

In the experimentally relevant regime $mR\gg1$ and $\beta\ll1$, the dynamics can be captured accurately by the truncated Wigner approximation (TWA) \cite{2003PhRvA..68e3604P, 2010AnPhy.325.1790P}. TWA is a semiclassical approach in which quantum fluctuations are accounted for by averaging over stochastic initial conditions \cite{2001PhRvL..87u0404S, 2002JPhB...35.3599S}, while the subsequent time evolution itself is governed by the classical equations of motion. In our simulation, we set parameters in the experimental regime; the detailed numerical implementation and parameter choices can be found in Ref. \cite{2025arXiv251213901F}. Our primary observable is the local expectation value of the phase cosine at the trap centre,
$$C(t)=\langle\cos\varphi(t,0)\rangle\,,$$
which characterises the dynamics of the relative phase of the condensates \cite{2007Natur.449..324H, 2013PhRvL.110i0404D}. We consider a protocol in which the initial state for the nonzero modes is the ground state of the massless theory ($m=0$), while the initial condition for the zero mode,
$$\varphi_0=\frac{1}{2R}\int_{-R}^{R}dx\,\varphi(x)\,,$$
is sampled from a Gaussian distribution characterised by mean value $\langle\varphi_0\rangle$ and width $\sigma_0$, with the variance of the conjugate momentum saturating the Heisenberg uncertainty relation \cite{2025arXiv251213901F}.

\textit{Results.}---  
We now turn to the post-quench dynamics of the inhomogeneous sine-Gordon model. Throughout this work, we use the term phase locking to denote the tendency of the phase distribution to become increasingly concentrated around $\varphi=0$ \cite{2018PhRvL.120q3601P}. Note that while $C(t)$ provides a local measure of this effect, it should not be identified directly with the dynamics of the global zero mode.

Fig.~\ref{fig:dephasing_and_phaselocking} reveals two parametrically separated dynamical regimes. At early times, the inhomogeneous system exhibits the same universal dephasing as the homogeneous sine-Gordon model: trajectories for different trap sizes remain nearly indistinguishable from the homogeneous result \cite{2013PhRvL.110i0404D, 2016PhRvE..93f2101K}. At later times, however, the dynamics departs from the homogeneous prediction and develops a pronounced drift toward larger expectation values, signalling the onset of phase locking \cite{2018PhRvL.120q3601P}. This drift is absent in the homogeneous model, where $C(t)$ remains approximately constant after the initial dephasing stage \cite{2019PhRvA.100a3613H}. 

Increasing the length parameter $R$ leaves the early-time dynamics essentially unchanged but delays the onset of phase locking. Therefore, we can identify two distinct timescales: a microscopic dephasing time set by the Josephson frequency $m$, and a much longer relaxation time $\tau_{\rm rel}$ determined by the curvature of the density profile at the trap centre. The dynamics crosses over from universal dephasing to a distinct long-time regime. For the harmonic profile $\tau_{\rm rel}=R$, and therefore the two scales are separated by the large factor $m\tau_{\rm rel}=mR$.

For a fixed profile shape $\tau_{\rm rel}\propto R$, so rescaling time by $R$
should collapse the long-time dynamics. Indeed, after such a rescaling, the time evolutions evaluated for different system sizes scale onto a common curve over a broad time window (Fig.~\ref{fig:phaselockingScaled}). Since size and curvature are tied together for a given shape, the collapse alone does not identify the controlling scale. Instead, this can be disentangled by varying the profile while holding $R$ fixed (Figs.~\ref{fig:mixed} and~\ref{fig:loglogtau}). Deviations from the collapse appear after the light-cone time, which is necessary for signals reflected from the condensate edges to return to the centre. For the harmonic profile, the light-cone time is $t_*=\pi R$
[Eq.~(S.1)], coinciding with the maximum of $C(t)$ in Fig.~\ref{fig:phaselockingScaled}. 
By then the drift amplitude, $\propto 1/\cosh(t/\tau_{\rm rel})$ [Eq.~\eqref{eq:taurel}], has fallen to
$1/\cosh\pi\approx 0.09$, so about 90\% of the drift is complete before any
boundary signal returns. Therefore, the universal drift is an intermediate regime between dephasing and the onset of finite-size effects \cite{Rauer2018,2021NatPh..17..559S}. Beyond $t_*$, the returning excitation causes moderate dips in $C(t)$, and $C(t)$ settles at $0.87$--$0.89$, far above the homogeneous plateau (Fig.~\ref{fig:phaselocking}). In the linearised theory, by contrast, $\langle\varphi(t,0)\rangle$ revives almost fully (Fig.~S.5), so the
persistence of the locked state can be attributed to the non-linearity already present in the sine-Gordon model.

In the homogeneous model, $C(t)$ dephases and stays on a long-lived prethermal plateau
\cite{2013NJPh...15g5011A}. In contrast, in the trap this prethermal plateau is short-lived, and the system enters a drift towards a phase-locked state, with $C(t)$ leaving the plateau on the timescale $\tau_{\rm rel}$  determined by the curvature (Figs. \ref{fig:dephasing_and_phaselocking} and \ref{fig:phaselockingScaled}).

To fully ascertain the origin of the drift, we compare the full dynamics with the linearised theory, $\cos\varphi\simeq1-\varphi^2/2$ (Fig.~\ref{fig:vslin}). The linearised model fails to reproduce the damping of the Josephson oscillations but captures the slow drift towards phase locking. Since the linearised theory is quadratic, the drift requires neither non-linear relaxation nor integrability breaking; it is already encoded in the mode structure generated by the inhomogeneous background. For $mR\gg1$ the linear theory can be solved asymptotically \cite{SM}: near a non-degenerate maximum of $f$, the mode operator reduces to an inverted harmonic oscillator, and the classical one-point function at the centre becomes
\begin{equation}
  \langle\varphi(t,0)\rangle \simeq
  \varphi_0\,\frac{\cos\!\big(m\sqrt{f(0)}\,t\big)}{\sqrt{\cosh(t/\tau_{\rm rel})}},
  \;
  \frac{1}{\tau_{\rm rel}}=\frac{1}{R}\sqrt{-\frac{f''(0)}{2}} .
  \label{eq:taurel}
\end{equation}
The fluctuations $\langle\varphi^2(t,0)\rangle$ decay as $1/\cosh(t/\tau_{\rm rel})$ resulting in $C(t)$ drifting upward. In the sine-Gordon model, the non-linearity also dephases the fast Josephson oscillations; hence $\langle\varphi\rangle$
itself deviates from Eq.~\eqref{eq:taurel} (Fig.~S.3). However, the slow drift
of $C(t)$ is still controlled by $\tau_{\rm rel}$ (Fig.~4). Since Eq.~\eqref{eq:taurel} is derived by replacing the finite trap with the whole real line, where recurrences are absent by construction, it is clear that the drift is set by the curvature of the density profile at its maximum, and not the finite-size effects responsible for revivals. In physical units, $\tau_{\rm rel}^{-1}=\sqrt{-g_{\rm 1D}\,\rho_0''(x_*)/2M}$ [Eq.~(S.44)], independent of the system size and of the tunnelling strength as observed for the damping time in \cite{2018PhRvL.120q3601P}. For a harmonically trapped condensate $\tau_{\rm rel}=\sqrt{2}/\omega_\parallel$, where $\omega_\parallel$ is the longitudinal trap frequency.

Curvature effects can be separated from finite-size effects by comparing different density profiles at fixed $R$. Fig.~\ref{fig:mixed} compares homogeneous ($f(z)=1$), quadratic ($f(z)=1-z^2$), arc ($f(z)=\sqrt{1-z^2}$) and ramp ($f(z)=1-z$), marking the light-cone time $t_*$ necessary for the first return of signals from the nearest boundary [Eq.~\eqref{eq:lightconetime}]. In the window $t<t_*$, the profiles without central curvature (homogeneous and ramp) show no drift, whereas both curved profiles do; the quadratic one drifts faster since its central curvature is twice that of the arc at the same $R$. The ramp curve rises only after its $t_*$, when the signal from the dense edge at $z=-1$ reaches the centre. 

These results show that the drift to phase locking is a curvature effect. Fitting Eq.~\eqref{eq:fitfunction} to the sine-Gordon data yields a relaxation time $\tau$ that follows Eq.~\eqref{eq:taurel} without adjustable parameters over more than a decade in $\epsilon$, with fitted exponents $-0.97(2)$ and $-0.94(2)$ for the quadratic and arc profiles (Fig.~\ref{fig:loglogtau}). In particular, Eq.~\eqref{eq:taurel} predicts, with no adjustable parameters, that at equal $R$ the arc relaxes more slowly than the quadratic profile by a factor of $\sqrt{2}$. Indeed, when time is measured in units of the analytic $\tau_{\rm rel}$, the two profiles collapse onto a single curve for $mR=25.5$, 51 and 102, increasingly well as $mR$ grows, until the arc reaches its light-cone time $t_*\approx 1.7\,\tau_{\rm rel}$ (Fig.~S.4 in \cite{SM}).

We mention that phase locking is also sensitive to the sine-Gordon parameter $\beta$ \cite{SM}: quantum fluctuations suppress it, and it disappears completely for
sufficiently large $\beta$, making it a semiclassical phenomenon.

\textit{Discussion.}---
We have identified a universal drift towards phase locking in the
post-quench dynamics of the inhomogeneous sine-Gordon model. At short times, the trapped system follows the universal dephasing of the homogeneous theory; at parametrically later times, the inhomogeneity induced by the slowly varying trap drives the central region of the condensate towards a phase-locked state. The time scale of the drift is controlled by the curvature of the density profile at its maximum through $\tau_{\rm rel}$ [Eq.~\eqref{eq:taurel}]. Since $\tau_{\rm rel}\propto R$ for a given
profile shape, this yields both the scaling collapse in $t/R$ and the parametric
separation $m\tau_{\rm rel}\sim mR$ from the Josephson dephasing time. Universality
thus means that the long-time drift is governed by a single local scale, independent
of the global shape of the trap. The universal drift regime occupies an intermediate temporal window between homogeneous dephasing and finite-size effects; beyond this window, non-linearity keeps the system locked by suppressing revivals. The two phenomena are nevertheless physically distinct: phase locking arises locally near the density maximum, whereas revivals result from finite-size effects.

A central insight is that the linearised theory already captures the drift
qualitatively, and that an asymptotic analysis of its collective modes yields the
timescale. The mechanism is therefore not primarily non-linear thermal relaxation.
Instead, the trap reshapes the collective modes of the relative phase: near the
density maximum they are governed by an inverted harmonic oscillator, which carries
the coherent excitation away from the centre and produces the slow drift of $C(t)$.
Non-linearity remains essential for damping the fast Josephson oscillations, but the
drift itself is encoded in the inhomogeneous collective-mode structure.

The mechanism we uncovered offers a natural interpretation of the phase locking observed in tunnel-coupled one-dimensional condensates, which homogeneous sine-Gordon dynamics alone cannot explain. More broadly, our analysis shows that slowly varying spatial inhomogeneity, often treated as a secondary experimental imperfection, can control the long-time dynamics of near-integrable quantum fields. Trapped condensates thus
provide a controlled setting for studying universal dynamics beyond homogeneous integrability, motivating future work on boundaries, trap engineering and quantum corrections beyond the semiclassical regime.

\paragraph*{Acknowledgments -}
\begin{acknowledgments}
We are grateful to J. Schmiedmayer for inspiring discussions. This work was supported by the HUN-REN Hungarian Research Network through the Supported Research Groups Programme, HUN-REN-BME-BCE Quantum Technology Research Group (TKCS-2024/34), and also by the National Research, Development and Innovation Office of Hungary (NKFIH) through OTKA Grant No. ANN 142584. B.F. was also partially supported by the Doctoral Excellence Fellowship Programme (DCEP), funded by the National Research, Development and Innovation Fund of the Ministry of Culture and Innovation and the Budapest University of Technology and Economics. 
ChatGPT 5.6 Sol contributed substantially to the derivation of the asymptotic formula; all results were verified by the authors.
\end{acknowledgments}

\bibliographystyle{utphys}
\bibliography{crossover_refs}

\clearpage
\onecolumngrid
\clearpage
\appendix
\setcounter{section}{0}
\setcounter{secnumdepth}{2}
\setcounter{equation}{0}
\setcounter{figure}{0}
\setcounter{table}{0}
\setcounter{theorem}{0}
\counterwithout{equation}{section}
\counterwithout{figure}{section}
\counterwithout{table}{section}
\counterwithout{equation}{section}
\renewcommand{\theequation}{S.\arabic{equation}}
\renewcommand{\thefigure}{S.\arabic{figure}}
\renewcommand{\thetable}{S.\arabic{table}}
\renewcommand{\thetheorem}{S.\arabic{theorem}}

\begin{center}
    \emph{Supplemental Material}
\end{center}

\section{Dependence on model parameters}\label{app:param}

\subsection{Dependence on quantum fluctuations}\label{subsec:quantum_fluctuations}

\begin{figure} [b]
    \centering
    \includegraphics{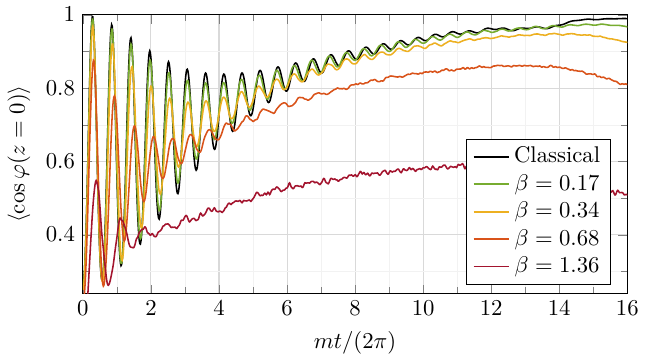}
    \caption{The expectation value $\langle\cos\varphi\rangle$ in the sine--Gordon model at different values of $\beta$. Parameters: $Rm = 25.5, \varphi_0 = -1.3, \sigma_{0} = 0.2$.}
    \label{fig:betadep}
\end{figure}

To assess the role of quantum fluctuations, we vary the parameter $\beta$, which controls the distance from the classical limit. Figure~\ref{fig:betadep} shows the expectation value of the cosine of the relative phase $C(t)=\langle\cos\varphi(t,0)\rangle$ for fixed initial conditions and different values of $\beta$.

We find that the phase-locking mechanism remains present throughout the semiclassical regime, but its magnitude decreases with increasing $\beta$ \cite{2019PhRvA.100a3613H, Szasz-Schagrin2022}. In the classical limit, $C(t)$ approaches values close to unity before revivals set in, whereas stronger quantum fluctuations progressively suppress the phase-locking drift \cite{2020PhRvB.101v4102K}. In the experimentally relevant regime, $\beta\simeq0.34$, the dynamics remains close to the classical trajectory, though it only reaches a peak value of $C(t) \approx 0.95$ \cite{2025arXiv251213901F}. This highlights that even in this deeply semiclassical regime, quantum fluctuations still play a discernible role in limiting the efficiency of the phase locking. For sufficiently large $\beta$, robust phase locking disappears altogether. In this deep quantum regime, the Truncated Wigner Approximation (TWA) eventually loses its validity. While the TWA remains an admissible approximation up to $\beta = 1.42$ \cite{2019PhRvA.100a3613H}, alternative approaches such as the Truncated Conformal Space Approach (TCSA) \cite{Yurov_1989,Feverati_1998,Horvath2022} are better suited for simulating the sine-Gordon model at these larger values of $\beta$.

These results indicate that the phase-locking mechanism itself is not of quantum origin. Rather, it is already present in the classical \cite{2023ScPP...15..140K} and linearised dynamics \cite{2013PhRvL.110i0404D} and originates from the collective-mode structure induced by the inhomogeneous background. Quantum fluctuations primarily determine the quantitative robustness of the phase-locked regime \cite{2003PhRvA..68e3604P}. In the semiclassical regime, the deviation of the peak cosine expectation value from unity scales approximately as $\beta^2$, consistent with quantum fluctuations providing the leading correction to the classical trajectory \cite{2003PhRvA..68e3604P, 2010AnPhy.325.1790P}; a more detailed derivation of this scaling is presented in Appendix \ref{subsec:classical_limit}.

We emphasise that interpreting such non-equilibrium results requires care, and that this reduction in the cosine expectation value should not be conflated with a standard thermal-equilibrium effect. Because the initial quench injects a finite energy density into the system, the asymptotic relaxation value of $C(t)$ would naturally deviate from unity even in the purely classical case due to an effective finite temperature. However, within the time window considered here, the system remains far from relaxation and strongly out of equilibrium \cite{2006PhRvL..96m6801C, 2012Sci...337.1318G}, precluding a straightforward equilibrium characterisation.

\subsection{Initial phase}\label{subsec:initial_phase}
\begin{figure}
    \centering
    \includegraphics{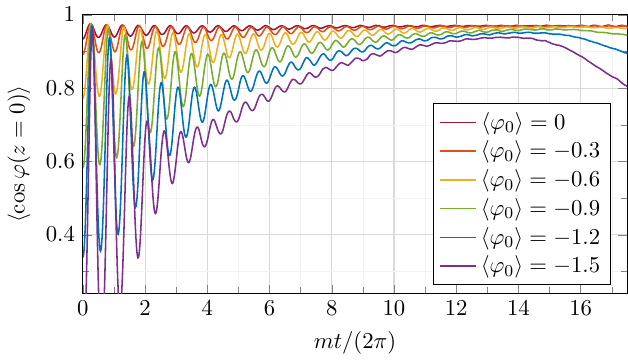}
    \caption{Comparison between simulations with different initial $\langle \varphi_0 \rangle$ expectation values ($0.,...,-1.5$). The graph shows the cosine expectation value of the relative phase as a function of time. Parameters: $mR = 25.5$, $\beta = 0.343$, and $\sigma_0 = 0.2$.}
    \label{fig:Phi0Dep}
\end{figure}

Starting from different initial zero-mode phases, the time evolution of the cosine expectation value is illustrated in Fig.~\ref{fig:Phi0Dep}. We observe that phase locking is robust to variations in the initial phase, and the corresponding time is essentially independent of it. When the initial phase is of order 1, the effects of non-linearity are enhanced and visible dephasing is observed; when the relative phase approaches zero, the dephasing weakens and then disappears.

\subsection{The $\langle\varphi\rangle$ expectation value}\label{subsec:phi_expval} 
\begin{figure}
    \centering
    \includegraphics{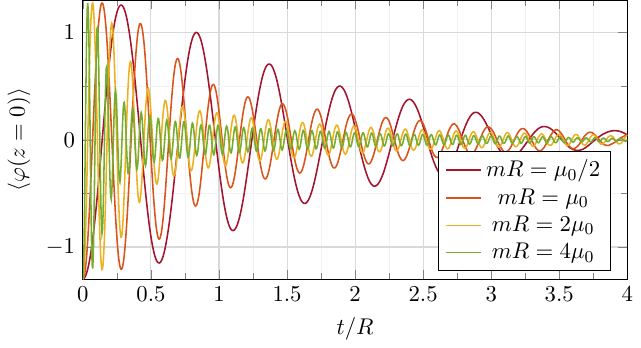}
    \caption{The expectation value $\langle\varphi\rangle$ at various system sizes. The simulation underlying this plot is identical to that in Fig.~\ref{fig:phaselockingScaled}; however, we do not observe the same scaling collapse. Parameters: $\mu = 25.5$, $\beta = 0.343$, and $\sigma_0 = 0.2$.}
    \label{fig:VolPhi}
\end{figure}

While the timescale of phase locking is determined solely by the local curvature of the density profile, this is not the case for other measurable quantities. The relaxation time depends on the expectation value considered in the measurement. While varying $mR$ leads to a scaling collapse of the $\langle\cos\varphi\rangle$ expectation value measured in units of $t/R$ (see Fig.~\ref{fig:phaselockingScaled}), we do not observe similar behaviour for the expectation value $\langle\varphi\rangle$ depicted in Fig.~\ref{fig:VolPhi}. In contrast to the phase locking time, the relaxation time of the expectation value $\langle\varphi\rangle$ scales differently. This behaviour is different from the results of the linearised model, which predicts that $\tau_{\langle\varphi\rangle}=2\tau_{\langle\cos\varphi\rangle}$. Therefore, non-linearity plays a crucial role in the relaxation process of $\langle\varphi\rangle$.

\subsection{The light-cone time}\label{subsec:light_cone_time}
The round-trip time from the centre to the nearest boundary can be obtained from integrating the position-dependent speed of sound $c(x)$ as
\begin{align}\label{eq:lightconetime}
    t_* = 2\min_{\pm}\int_0^{\pm R}\frac{dx}{c(x)} = \frac{2R}{c_0}\,\min_{\pm}\left|\int_0^{\pm 1}\frac{dz}{\sqrt{f(z)}}\right|\,.
\end{align}
Note that it is a lower bound, since massive modes travel slower than $c(x)$, and the reflection from the boundary also entails a further time shift.

\subsection{The fitted formula}\label{subsec:fitted_formula} To extract the relaxation time $\tau$, we fit the following function with three parameters $(A,\tau,\delta)$ 
\begin{align} \label{eq:fitfunction}
    C_\mathrm{fit}(t|A,\tau,\delta) = 1 - A \frac{1/2}{\cosh(t/\tau)}+\delta\,.
\end{align}
The exact form is based on the asymptotic formula Eq.~\eqref{eq:twoPointFormula}. As we vary the curvature, we observe a relatively small change in the set of parameters $\{A,\delta\}$, confirming the robustness of our approach. Note that the fit is applied to the cosine of the phase while the fit function itself comes from a linear approximation, which entails some errors. Moreover, the fitted parameters exhibit residual dependence on the fit window. The ratio of the relaxation times plotted in Fig.~\ref{fig:loglogtau} recovers the linear predictions with an error of $10-15\%$. A better qualitative demonstration of the analytic result can be seen by an explicit comparison of the rescaled curves.

\begin{figure}[t]
    \centering
    \subfloat[$mR=25.5$]{%
    \includegraphics{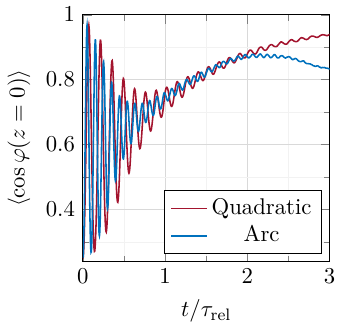}
    }
    \subfloat[$mR=51$]{%
    \includegraphics{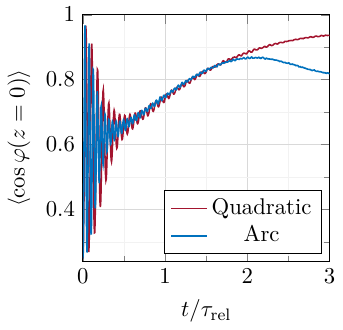}
    }
    \subfloat[$mR=102$]{%
    \includegraphics{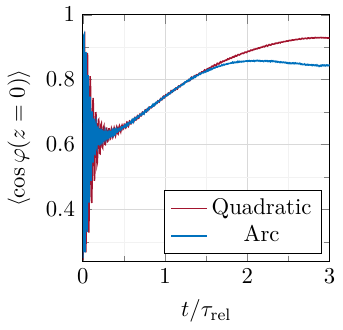}
    }
    \caption{The cosine of the relative phase at the centre of the condensate as a function of the time measured in units of $\tau_\mathrm{rel}$ (calculated with the analytic formula cf.~Eq.~\eqref{eq:taurel}) with the quadratic and arc profiles. The different plots showcase the same data with different system sizes. We observe that the two sets of data collapse into a single curve on the timescale of $\tau_\mathrm{rel}$. Parameters: $R m = 25.5, 51, 102, \beta = 0.343, \langle\varphi_0\rangle = -1.3, \sigma_0=0.2$.} \label{fig:MixedScaled}
\end{figure}

\paragraph*{Scaling collapse with different profiles -} As a more direct way to probe the validity of the analytic formula Eq.~\eqref{eq:taurel}, we demonstrate the scaling collapse of the coherence measured as a function of the time rescaled by $\tau_\mathrm{rel}$ in Fig.~\ref{fig:MixedScaled}. Here we determine $\tau_\mathrm{rel}$ analytically, based on Eq.~\eqref{eq:taurel}, i.e., there is an exact $\sqrt{2}$ relative scaling factor in physical time between the two curves. We observe that the two curves collapse on a wide range at the scale of $\tau_\mathrm{rel}$.

\newpage
\section{Dynamics of the linearised model}\label{app:lin}

This appendix summarises the analytic results of the linearised model and cross-checks them against the TWA simulations.

For simplicity, we introduce the dimensionless variable $\mu = mR$, and the dimensionless space-like coordinate $z = x/R$, which is an element of the interval $(-1,1)$. We can rewrite the action \eqref{eq:actionDim} in terms of dimensionless variables as
\begin{align}\label{eq:linS}
    S = \int d^2 x \left( \frac12 (\partial_t \phi)^2 - \frac12 f(z) (\partial_z\phi)^2 - \frac{\mu^2}{2}f(z)\,\phi^2 \right)\,,
\end{align}
where $t$ is measured in units of $R$.

\subsection{Diagonalizing the model}

The Hamiltonian can be diagonalised by expanding the field in eigenmodes, which are the solutions of the classical equation of motion
\begin{align}
    -\partial_t^2\phi_\textrm{cl} + \partial_z(f\cdot\partial_z\phi_\textrm{cl})-\mu^2 f(z)\phi_\textrm{cl} = 0\,,
    \label{eq:classical_linear_eqn_of_motion}
\end{align}
and (from the boundary terms in the partial integration in $\delta S = 0$) the boundary condition reads as
\begin{align}
    \lim_{z\rightarrow\pm 1} f(z) \partial_z\phi(t,z) = 0\,.
\end{align}
The above partial differential equation can be solved by the usual method of separating the variables $\phi(t,z) = T(t) Z(z)$. The equation
\begin{align*}
    -\frac{T''}{T} +\frac{(f Z')'-f \mu^2 Z}Z = 0\,,
\end{align*}
can be separated by introducing the energy $\omega$
\begin{align*}
    -\frac{T''}{T} = \omega^2\,,
\end{align*}
which is determined by the eigenvalue equation
\begin{align*}
    (f Z')'-f \mu^2 Z = -\omega^2 Z
\end{align*}
with the boundary condition mentioned above. The spectrum is discrete and bounded from below, so we enumerate the energies and eigenvectors as $\omega_n^\mu, Z_n^\mu$ \footnote{Here we indicated the $\mu$ dependence as we will consider quenches in $\mu$. Although the eigenvalues and eigenvectors also depend on the profile, we suppress this dependence to simplify the notation.}. Then, if we consider the mode expansion of the field $\phi(t,z)$ in this basis
\begin{align*}
    \phi(t,z) = \sum_n Z_n^\mu(z) \phi_n^\mu(t) =\sum_n \frac{Z_n^\mu(z)}{\sqrt{2\omega_n^\mu}} \left(a_n^\mu e^{-i\omega_n^\mu t} + {a_n^\mu}^\dagger e^{i\omega_n^\mu t} \right)\,,
\end{align*}
the Hamiltonian becomes diagonal
\begin{align} \label{Eq:diagH}
    H = \int_{-1}^1 dz \left(\frac12 (\partial_t \phi)^2 + \frac12 f(z) (\partial_z\phi)^2 + \frac{\mu^2}{2}f(z)\,\phi^2\right) = \sum_n \omega_n^\mu {a_n^\mu}^\dagger a_n^\mu + \textrm{const.}\,.
\end{align}
Note that the $Z_n^\mu$ functions form a complete system, and we chose them to be orthonormal
\begin{align*}
    \int_{-1}^1 dz Z_n^\mu(z)Z_m^\mu(z) = \delta_{nm}\,.
\end{align*}
For a massless theory, the zero mode must be handled separately:
\begin{align}
    \phi(t,z) = \phi_0 Z_0^0(z) + \Pi_0 Z_0^0(z) \,t + \sum_{n\neq 0} \frac{Z_n^0(z)}{\sqrt{2\omega_n^0}} \left(a_n^0 e^{-i\omega_n^0 t} + {a_n^0}^\dagger e^{i\omega_n^0 t} \right)\,,
\end{align}
where $Z_0^0(z) \equiv 1/\sqrt2$. The Hamiltonian retains the same form as in Eq.~\eqref{Eq:diagH}, where the zero mode has a separate contribution of $\Pi_0^2/2$ (of which the excitation spectrum is manifestly different from the higher modes). For the quench protocols considered here, the initial state is the ground state of the higher modes, while the zero-mode part is specified separately. Defining the zero mode in the $\phi_0$ eigenbasis, we consider Gaussian squeezed states of the form
\begin{align*}
    \psi_{\langle\phi_0\rangle,\sigma}(\varphi_0) = \left(\sqrt{\pi} \,\sigma\right)^{-1/2} \exp\left(-\frac{(\varphi_0-\langle\phi_0\rangle)^2}{2\sigma^2}\right)
\end{align*}
parametrised by a single parameter $\sigma$. This, in fact, means that the zero-mode wave function has variances
\begin{align*}
    \sigma_{\phi_0}^2 &= \frac{\sigma^2} 2\qquad\text{and}\\
    \sigma_{\Pi_0}^2 &= \frac{1}{2\sigma^2}\,.
\end{align*}
Note that, for a mixed zero-mode state (e.g., a finite-temperature state), the variances $\sigma_{\phi_0}^2$ and $\sigma_{\Pi_0}^2$ may be treated as independent, provided they satisfy the Heisenberg uncertainty relation.

The proper matching between $\langle\phi_0\rangle, \sigma$ and the experimental parameter, initial total phase difference of the condensates, $\varphi_0$, is given by
\begin{align}
\begin{aligned} \label{Eq:initParam}
    \langle\phi_0\rangle &= \frac{\sqrt2}\beta  \langle \varphi_{0}\rangle\,,\\
    \sigma &= \sqrt 2 \sigma_{\phi_0} = \frac{2}{\beta} \sigma_{0}\,.
\end{aligned}
\end{align}
Note that the factor of $\sqrt 2$ originates solely from the difference in normalisation conventions for the experimental zero mode (denoted by the subscript $0$).

\subsection{Time evolution of a Bogoliubov quench}

We now consider the standard Bogoliubov quench protocol, i.e., we consider a scenario when the mass is changed as $\mu_\mathrm{pre} \rightarrow \mu$. We focus on the case $\mu_\mathrm{pre} = 0$, although the more general case can be treated analogously.

Initially, the system is prepared in the ground state of a massless theory, with the zero mode described by the squeezed state introduced above. We consider the time evolution of this state in the post-quench theory, where we add a mass term parametrised by $\mu$. The pre-quench state $|0\rangle$ is described by the zero mode wave function, while the higher modes are characterised by annihilation from the pre-quench operators: $a_n^0|0\rangle = 0 = \langle 0|a_n^0$ for all $n > 0$. It therefore remains to determine the relation between the pre-quench and post-quench creation and annihilation operators. Writing the field $\phi$ and $\Pi \equiv \partial_t \phi$ in terms of the two different mode expansions
\begin{align*}
    \begin{aligned}
    \phi(0,z) &= \phi_0 Z_0^0 + \sum_{n\neq 0} \frac{Z_n^0(z)}{\sqrt{2\omega_n^0}} \left(a_n^0 + {a_n^0}^\dagger\right)= \sum_{n} \frac{Z_n^\mu(z)}{\sqrt{2\omega_n^\mu}} \left(a_n^\mu + {a_n^\mu}^\dagger\right)\,,
    \\
    \Pi(0,z) &= \Pi_0 Z_0^0 + \sum_{n\neq 0} Z_n^0(z)(-i)\sqrt\frac{\omega_n^0}{2} \left(a_n^0 - {a_n^0}^\dagger\right)= \sum_{n} Z_n^\mu(z)(-i)\sqrt\frac{\omega_n^\mu}{2} \left(a_n^\mu - {a_n^\mu}^\dagger\right)\,,
    \end{aligned}
\end{align*}
one can express
\begin{align}
    \begin{aligned}
        a_n^\mu = U_{n0}^\mu \sqrt{\frac{\omega_n^\mu}{2}}\left(\phi_0 + \frac{i}{\omega_n^\mu}\Pi_0\right) + \sum_{m\neq 0} U_{nm}^\mu \left(\sqrt{\frac{\omega_n^\mu}{\omega_m^0}}\frac{a_n^0+{a_n^0}^\dagger}{2}+\sqrt{\frac{\omega_m^0}{\omega_n^\mu}}\frac{a_n^0-{a_n^0}^\dagger}2\right)\,,
    \end{aligned}
\end{align}
where the transformation matrix is
\begin{align*}
    U_{nm}^\mu = \int_{-1}^1 dz Z_n^\mu(z) Z_m^0(z)\,.
\end{align*}
The time evolution can be expressed in terms of the post-quench mode expansion, with the pre-quench creation and annihilation operators replaced by their post-quench counterparts. Here, we consider the expectation values of the eigenmodes of the post-quench Hamiltonian:
\begin{align*}
    \phi_n^\mu(t) = \int_{-1}^1 dz Z_n^\mu(z) \phi(t,z)\,.
\end{align*}
After substitution, exploiting the properties of creation and annihilation operators, their relation to the pre-quench vacuum and using the squeezed Gaussian form for the zero-mode profile, the one-point function of eigenmodes becomes
\begin{align*}
    \langle0|\phi_n^\mu(t)|0\rangle \equiv \langle\phi_n^\mu(t)\rangle &= \langle\phi_0\rangle U_{n0}^\mu \cos(\omega_n^\mu t)\,.
\end{align*}
Similarly, the connected two-point function of eigenmodes,
\begin{align*}
    \langle\phi_n^\mu(t)\phi_m^\mu(t)\rangle_c &= \langle0|\phi_n^\mu(t)\phi_m^\mu(t)|0\rangle - \langle\phi_n^\mu(t)\rangle\langle\phi_m^\mu(t)\rangle\,,
\end{align*}
gives
\begin{align}
    \begin{aligned} \label{eq:phiphinmc}
        \langle\phi_n^\mu(t)\phi_m^\mu(t)\rangle_c &= \frac{\sigma^2}{2}U_{n0}^\mu U_{m0}^\mu\left(\cos(\omega_n^\mu t) - \frac{i}{\sigma^2\omega_n^\mu}\sin(\omega_n^\mu t)\right)\left(\cos(\omega_m^\mu t) + \frac{i}{\sigma^2\omega_m^\mu}\sin(\omega_m^\mu t)\right)\\
        &+\frac{1}{2}\sum_{k\neq 0}U_{nk}^\mu U_{mk}^\mu\frac{1}{\omega_k^0}\left(\cos(\omega_n^\mu t) - i\frac{\omega_k^0}{\omega_n^\mu}\sin(\omega_n^\mu t)\right)\left(\cos(\omega_m^\mu t) + i\frac{\omega_k^0}{\omega_m^\mu}\sin(\omega_m^\mu t)\right)\,.
    \end{aligned}
\end{align}
The real-space one- and two-point functions become
\begin{align}
    \langle\phi(t,z)\rangle &= \sum_{n} Z_n^\mu(z) \langle\phi_n^\mu(t)\rangle\,,\nonumber\\
    \langle\phi(t,z)\phi(t,z')\rangle &= \sum_{n,m} Z_n^\mu(z) Z_m^\mu(z') \langle\phi_n^\mu(t)\phi_m^\mu(t)\rangle\,.
\end{align}

To compare the analytic results for $\langle\phi(t,z)^2\rangle$ to the TWA simulations, the field normalisations and regularisations must be matched. As $z \rightarrow z'$, the two-point function has a logarithmic singularity (cf.~free boson CFT). We define the renormalised $\phi(t,z)^2$ field as the $z\rightarrow z'$ limit of the regular part of the $\phi(t,z)\phi(t,z')$ OPE. With our normalisation convention of the action, the normal ordering becomes
\begin{align}
    :\phi(t,z)^2: &= \left(\frac1a\int_{z-a/2}^{z+a/2} dz' \phi(t,z')\right)^2 + \frac{1}{2 \pi}\log a + \textrm{reg.}\,.
\end{align}
On a lattice with lattice spacing $a$, it reads as
\begin{align}
    :\phi_{z_i}^2: &= \phi_{z_i}^2 + \frac{1}{2\pi} \log a + \textrm{const.}\,.
\end{align}
We keep the lattice spacing constant throughout all the TWA simulations.

\subsection{Classical limit}\label{subsec:classical_limit}

We now consider the limit $\beta\rightarrow 0$, corresponding to the semiclassical regime. We consider the field $\varphi = \beta \phi$, which is kept finite in this limit. While $\beta$ is not a parameter of the linear model, it appears in the input variables $\langle\phi_0\rangle$ and $\sigma$. Having the zero mode parameters of the field $\varphi$ fixed, the only $\beta$-dependent part of the multi-point functions of $\phi$ is the zero mode. Consequently, in this limit, the non-zero-mode contributions are projected out:
\begin{align} \label{eq:clvarphi}
    \lim_{\beta\rightarrow 0}\langle\varphi(t, z)\rangle &= \lim_{\beta\rightarrow 0}\beta \langle\phi(t, z)\rangle = \sqrt2 \varphi_{0}\sum_n Z_n^\mu(z) U_{n0}^\mu \cos(\omega_n^\mu t)\,,\nonumber\\
    \lim_{\beta\rightarrow 0} \langle\varphi(t,z)\varphi(t,z')\rangle_c &= (\sqrt{2}\sigma_{0})^2 \sum_{n,m} Z_n^\mu(z) Z_m^\mu(z') U_{n0}^\mu U_{m0}^\mu \cos(\omega_n^\mu t)\cos(\omega_m^\mu t)\,,
\end{align}
which is nothing but the solution of the classical equation of motion \eqref{eq:classical_linear_eqn_of_motion} obtained with the method of separation of variables. This limit is reached when the contribution of the higher modes in Eq.~\eqref{eq:phiphinmc} is negligible. Contributions of higher modes encode the effects of quantum fluctuations, which are suppressed in the classical limit. When these terms become large, i.e., much larger than the contribution of the zero mode, the $\langle\phi(t,z)\phi(t,z')\rangle_c$ correlator becomes independent of $\beta$, and therefore, the $\langle\varphi(t,z)\varphi(t,z')\rangle_c$ correlator is proportional to $\beta^2$, which is observed in Fig.~\ref{fig:betadep}.

Note that the one-point function of $\varphi$ is completely independent of $\beta$, that is, 
\begin{align}
    \lim_{\beta\rightarrow 0}\beta\langle\phi(t, z)\rangle = \beta\langle\phi(t, z)\rangle
\end{align}
for any $\beta$ (within the framework of the linear model). The contribution of quantum fluctuations is manifested only in the two-point function.

In the classical limit, the following relation holds between the one- and two-point functions
\begin{align}
    \lim_{\beta\rightarrow 0}\frac{\langle\varphi(t, z)\varphi(t, z')\rangle_c}{(\sqrt 2 \sigma_{0})^2} = \frac{\langle\varphi(t, z)\rangle}{\sqrt 2 \varphi_{0}} \frac{\langle\varphi(t, z')\rangle}{\sqrt 2 \varphi_{0}}\,.
\end{align}

As phase locking was observed in the regime when $\mu$ is large and $\beta$ is small (semiclassical), we consider the large $\mu$ asymptotic form of the classical solution given by the following theorem.
\begin{theorem}\label{th:largemu} Consider the action Eq.~\eqref{eq:linS} and suppose $f(z)$ has a local maximum at $z = 0$. The one-point function of $\varphi=\beta \phi$ follows the $\mu\gg1$ asymptotic formula
\begin{align}
\begin{aligned} \label{eq:analformula}
     \langle\varphi(t, 0)\rangle &= \sqrt2 \varphi_{0}\sum_n Z_n^\mu(0) U_{n0}^\mu \cos(\omega_n^\mu t) \simeq \varphi_{0}\,\frac{\cos\left(\mu t\sqrt{f(0)}\right)}{\sqrt{\cosh\!\left(t/\tau_{\mathrm{rel}}\right)}}\,,
\end{aligned}
\end{align}
where
\begin{align}
    \frac{1}{\tau_{\mathrm{rel}}}
    =
    \sqrt{-\frac12 f''(0)}
\end{align}
which is valid for $t/R \lesssim \mathcal{O}(\log \mu)$.
\end{theorem}
Note that the time is measured in units of $R$.
\begin{proof}
The proof is relegated to Appendix \ref{app:proof}.
\end{proof}

With this formula, the two-point function has the following behaviour in the classical limit (using $f(0) = 1$)
\begin{align}
    \lim_{\beta\rightarrow 0} \langle\varphi(t,0)^2\rangle &\simeq (\sigma_{0}^2 + \varphi_{0}^2)\frac{\cos^2(\mu t)}{\cosh(t/\tau_{\mathrm{rel}})}\,.
\end{align}
The leading quantum correction can be approximated as a constant shift coming from the higher mode terms in Eq.~\eqref{eq:phiphinmc}. Combining with our previous result, the two-point function in the centre can be approximated as
\begin{align} \label{eq:twoPointFormula}
    \langle\varphi(t, 0)^2\rangle \simeq \left(\varphi_{0}^2 +\sigma_{0}^2\right) \frac{\cos^2(\mu t)}{\cosh(t/\tau_{\mathrm{rel}})}+\delta\,,
\end{align}
where $\delta$ is the approximate shift due to quantum fluctuations. Note that while $\delta$ is zero in the classical limit, it does not measure \emph{quantumness}: a finite initial temperature (of the higher modes) can also decrease the phase-locked expectation value.

\subsection{Cross-checks with numerics}

\begin{figure}
    \centering
    \includegraphics[width=0.48\linewidth]{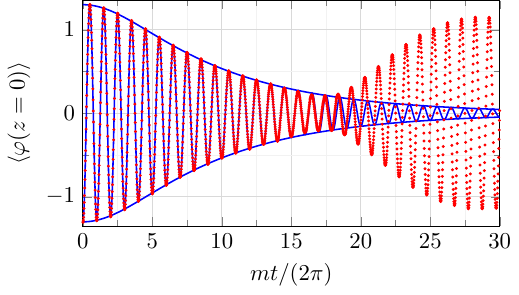}
    \includegraphics[width=0.48\linewidth]{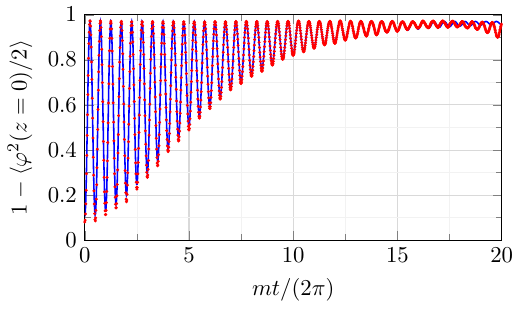}
    \caption{Comparison between the TWA simulation of the linear model and the analytic approximation. The left panel shows the relative phase at the centre of the condensate. The red dots represent the TWA results, while the blue curves denote the analytic prediction of Eq.~\eqref{eq:analformula} together with its upper and lower envelopes. The right panel shows the expectation value of $\cos\varphi$ at the centre of the condensate, evaluated within the approximation $\cos\varphi \approx 1 - \varphi^2/2$. The red dots correspond to the TWA results, whereas the blue curves represent the analytic prediction for the two-point function (Eq.~\ref{eq:twoPointFormula}), including the leading quantum correction parameterised by $\delta \approx 0.03$. Parameters: $\mu = 25.5$, $\beta = 0.343$, $\langle \varphi_0 \rangle = -1.3$, and $\sigma_0 = 0.2$. Overall, the analytic approximation captures both the oscillatory dynamics of the relative phase and the evolution of the corresponding correlation function with good quantitative accuracy.}
    \label{fig:PhiVsAnal}
\end{figure}

Figure \ref{fig:PhiVsAnal} compares the analytic predictions with the TWA simulations. We observe that the expectation value of the phase at the centre of the condensate agrees remarkably well with the analytic prediction of Eq.~\eqref{eq:analformula} until the revivals appear. We also consider the second-order approximation for the expectation value of $\cos\phi$. After fitting the $\delta$ shift (cf.~Eq.~\eqref{eq:twoPointFormula}), we again observe remarkable agreement between the analytic and numerical results. 
In the experimentally relevant parameter regime, the fitted shift, $\delta \approx 0.03$, is small compared with the coherent term; its magnitude reflects the zero-point fluctuations of the coupled ground state at the lattice resolution.

\newpage
\section{Local large-\texorpdfstring{$\mu$}{mu} relaxation close to a density extremum}
\label{app:proof}

We consider the dimensionful quadratic action
\begin{align}
    S
    =
    \int dt\int_{-R}^{R} dx\,
    \left[
        \frac12(\partial_t\phi)^2
        -\frac12 f(x/R)(\partial_x\phi)^2
        -\frac{m^2}{2}f(x/R)\phi^2
    \right],
\end{align}
with \(\mu=mR\gg1\). Let \(z=x/R\), and suppose that \(f\) has a
non-degenerate local maximum at \(z=z_*\),
\begin{align}
    f'(z_*)=0,
    \qquad
    f''(z_*)<0,
    \qquad
    f_*:=f(z_*)>0.
\end{align}
We now evaluate the field at a nearby point \(z_0\), rather than directly at
the extremum.

Introducing the Liouville coordinate
\begin{align}
    \frac{dy}{dz}=\frac{1}{\sqrt{f(z)}},
\end{align}
and rescaling the modes by \(f^{1/4}\), the spatial operator takes
Schrödinger form. To leading order at large \(\mu\),
\begin{align}
    \widetilde{\mathcal L}
    \simeq
    -\partial_y^2+\mu^2 f(z(y)).
\end{align}
Expanding around the maximum,
\begin{align}
    f(z(y))
    =
    f_*-a(y-y_*)^2+\cdots,
    \qquad
    a=-\frac12 f_*f''(z_*)>0.
\end{align}
With
\begin{align}
    \eta
    =
    \sqrt{2\mu\sqrt a}\,(y-y_*),
    \qquad
    B=-\partial_\eta^2-\frac{\eta^2}{4},
\label{eq:rescaling_time}\end{align}
one obtains
\begin{align}
    \sqrt{\widetilde{\mathcal L}}
    \simeq
    \mu\sqrt{f_*}+\nu B,
    \qquad
    \nu=\sqrt{-\frac12 f''(z_*)}.
\end{align}

The observation point is characterised by the scaled displacement
\begin{align}
    \eta_0
    =
    \sqrt{2\mu\sqrt a}\,
    \bigl[y(z_0)-y(z_*)\bigr].
\end{align}
Equivalently, to quadratic order close to the extremum,
\begin{align}
    \frac{\eta_0^2}{4}
    \simeq
    \frac{\mu\bigl[\sqrt{f_*}-\sqrt{f(z_0)}\bigr]}{\nu},
\end{align}
so \(\eta_0\) measures the local frequency detuning from the extremum in units
of the relaxation rate.
For \(\eta_0=\mathcal O(1)\), corresponding to the natural
near-extremum semiclassical region, the quadratic approximation gives
\begin{align}
    \left(e^{-i\nu Bs}1\right)(\eta_0)
    =
    \frac{1}{\sqrt{\cosh(\nu s)}}
    \exp\left[
        \frac{i\eta_0^2}{4}\tanh(\nu s)
    \right],
\end{align}
where \(s=t/R\). Hence, for a locally homogeneous initial condition
\begin{align}
    \phi(x,0)=\phi_0,
    \qquad
    \partial_t\phi(x,0)=0,
\end{align}
the leading large-\(\mu\) result becomes
\begin{align}
    \boxed{
    \phi(x_0,t)
    \simeq
    \phi_0\,
    \frac{
        \cos\left[
            m\sqrt{f_*}\,t
            -\dfrac{\eta_0^2}{4}
             \tanh\!\left(\dfrac{t}{\tau_{\mathrm{rel}}}\right)
        \right]
    }{
        \sqrt{\cosh\!\left(t/\tau_{\mathrm{rel}}\right)}
    }
    },
\end{align}
with
\begin{align}
    \boxed{
    \frac{1}{\tau_{\mathrm{rel}}}
    =
    \frac{1}{R}\sqrt{-\frac12 f''(z_*)}
    }.
\end{align}
At the extremum, \(\eta_0=0\), and the result reduces to the expression
derived previously. Moving away from the extremum therefore leaves the
universal \(\cosh^{-1/2}\) envelope unchanged at leading order, while producing
a displacement-dependent phase shift governed by \(\tanh(t/\tau_{\mathrm{rel}})\).

The result is naturally interpreted as a crossover description close to the
extremum. The relevant condition is not \(f'(z_0)=0\) itself, but that the
observation point remains in the quadratic scaling region,
\begin{align}
    \eta_0=\mathcal O(1).
\end{align}
In this regime \(f'(z_0)\) is generally nonzero, but the local dynamics is still
controlled by the displaced inverted harmonic oscillator.

The derivation is local in two respects. First, we expanded the potential around
the top of the barrier at $z=0$. Second, after the rescaling \eqref{eq:rescaling_time}, we replaced the original finite interval by the whole
line. Therefore, the result should be interpreted as a large-$\mu$ approximation at
fixed times,
\begin{align*}
    t=\mathcal O(1),
    \qquad
    \mu\to\infty.
\end{align*}
We note that the approximation is not expected to be uniform at very large times. Indeed, the envelope predicted by the local approximation decays as
\begin{align*}
    \frac{1}{\sqrt{\cosh t}}
    \sim
    \sqrt{2}\,e^{-t/2}
    \qquad
    (t\gg1).
\end{align*}
At sufficiently late times, this exponentially small local contribution becomes
comparable to finite-$\mu$ corrections coming from the discreteness of the
spectrum, the finite interval, and the parts of the potential away from the
quadratic barrier approximation. These global effects are precisely what give rise to the revivals. A natural estimate for the breakdown time is therefore
logarithmic in $\mu$,
\begin{align*}
    t_{\mathrm{br}}
    \sim
    \mathcal O(\log\mu),
\end{align*}
up to non-universal constants depending on the observable and on how one defines
the onset of the revival.

\section{Connection to tunnel-coupled one-dimensional condensates}

The above asymptotic result has a direct interpretation for tunnel-coupled
one-dimensional condensates. In the weakly interacting, large-density regime,
the phase-density expansion of the Lieb--Liniger model around the
Gross--Pitaevskii profile gives an inhomogeneous Luttinger liquid. For two identical condensates each described by the Hamiltonian
\begin{align}
    H_{i} = \int dx\, \psi_{i}^\dagger(x) \left(-\frac{\hbar^2}{2M}\partial_x^2 + V(x) - \mu_c + \frac{g_\mathrm{1D}}{2}\psi_{i}^\dagger(x)\psi_{i}(x)\right)\psi_{i}(x)\,,\quad i=1,2\,,
\end{align}
which can be rewritten in terms of density-phase variables using $\psi_i=e^{i\theta_i}\sqrt{\rho_i}$. The tunnelling term
\(-\hbar J(\psi_1^\dagger\psi_2+\psi_2^\dagger\psi_1)\) generates a sine-Gordon potential for the
relative phase
\begin{align}
    \varphi=\theta_1-\theta_2\,,
\end{align}
which, in the linear approximation, results in the following equation of motion
\begin{align}
    \partial_t^2\varphi
    -
    c_0^2\partial_x\!\left[f(x/R)\partial_x\varphi\right]
    +
    c_0^2m_J^2 f(x/R)\varphi
    =
    0,
\end{align}
where
\begin{align}
    f(x/R) = \frac{\rho_0(x)}{\rho_0(0)},
\end{align}
$\rho_0$ is the classical density profile, $R$ is the Thomas--Fermi radius, $c_0 = c(0)$,
\begin{align}
    c^2(x)=\frac{g_{1\mathrm D}\rho_0(x)}{M}
\end{align}
the local sound velocity (squared), and
\begin{align}
    m_J^2=\frac{4MJ}{\hbar}
\end{align}
for $\hbar J \ll g_{1\mathrm D}\rho_0$ if \(J\) is measured as a frequency. After rescaling time by the reference sound
velocity, the theory is precisely of the form considered above, with
\(\mu=m_JR\) (for details we refer the reader to Ref. \cite{2025arXiv251213901F}). Close to a non-degenerate maximum \(x_*\) of the density profile, the relative phase therefore obeys
\begin{align}
    \boxed{
    \varphi(x_0,t)
    \simeq
    \varphi_0\,
    \frac{
        \cos\left[
            \omega_J(x_*)t
            -\dfrac{\eta_0^2}{4}
             \tanh\!\left(\dfrac{t}{\tau_{\mathrm{rel}}}\right)
        \right]
    }{
        \sqrt{\cosh\!\left(t/\tau_{\mathrm{rel}}\right)}
    }
    } ,
\end{align}
where
\begin{align}
    \omega_J(x_*)
    =
    \sqrt{
        \frac{4Jg_{1\mathrm D}\rho_0(x_*)}{\hbar}
    }
\end{align}
In physical terms, the displacement parameter can be written as
\begin{align}
    \frac{\eta_0^2}{4}
    \simeq
    \tau_{\mathrm{rel}}
    \bigl[\omega_J(x_*)-\omega_J(x_0)\bigr],
\end{align}
i.e.\ it measures the local Josephson-frequency detuning relative to the
relaxation rate.
The relaxation timescale reads as
\begin{align}
    \boxed{
    \frac{1}{\tau_{\mathrm{rel}}}
    =
    c(x_*)
    \sqrt{
        -\frac{\rho_0''(x_*)}{2\rho_0(x_*)}
    }
    =
    \sqrt{
        -\frac{g_{1\mathrm D}}{2M}\rho_0''(x_*)
    }
    }.
\end{align}
For a harmonically trapped condensate, $V(x) = \frac{1}{2}M\omega_{||}^2x^2$, and in the Thomas-Fermi regime this expression reduces to
\begin{align}
    \tau_{\rm rel}=\sqrt{2}/\omega_\parallel\,,
\end{align}
where $\omega_\parallel$ is the longitudinal trap frequency. 

The results of this section demonstrate that while the tunnelling strength fixes the fast Josephson oscillation, the time scale of phase locking is determined by the local curvature of the trapped density profile. The regulator \(R\) disappears from the physical relaxation time. 

\end{document}